\documentclass[leqno]{article}
\usepackage{amsmath,amssymb,amsthm,bm}
\usepackage{caption,graphicx}
\usepackage[all]{xy}
\theoremstyle{plain}

\def\be{\begin{equation}}
\def\ee{\end{equation}}

\title{On the multistationarity of chemical reaction networks}
\author{Marcelle {\sc Kaufman}$^a $\footnote{Corresponding author}  \ and Christophe {\sc Soul\'e}$^b$}
\date{$^a$ {\small Universit\'e libre de Bruxelles (ULB), Service de Chimie physique et Biologie th\'eorique,
Campus Plaine ULB, CP 231, Boulevard du Triomphe, 1050 Bruxelles, Belgique, mkaufman@ulb.ac.be }
\\ 
$^{b}$ {\small CNRS et IHES, Le Bois-Marie, 35 route de Chartres, 91440 {\sc Bures-sur-Yvette}, France, soule@ihes.fr}}

\begin{document}

\maketitle

\noindent We present a new conjecture about a necessary condition that a (bio)chemical network has to satisfy for it to exhibit  multistationarity. 
According to a Theorem of Feliu and Wiuf, the conjecture
is known for strictly monotonic
kinetics. We give several examples illustrating our conjecture.

\bigskip


\section{\bf  Introduction}


This paper is a contribution to the theory of multistationarity of (bio)chemical networks. The mathematical theory of chemical reaction networks is already well developed, thanks to the pioneering work of Feinberg \cite{F1,F2}, and major contributions by Craciun \cite{C1,GF,C}, Banaji \cite{BA}, Mincheva-Roussel \cite{M}, Wiuf and Feliu \cite{WF}, and others.

\smallskip

A {\em chemical reaction network} is a finite set of reactions between complexes made  of {\em chemical species}. These reactions depend upon a {\em stoichiometric matrix}, and to each of them is attached a {\em rate function} which controls the {\em kinetics} of the {\em concentration} value of the different species. For instance, in the classical case of mass-action kinetics, the rate functions are monomials in the concentration values.

\smallskip

One is interested in deciding whether such a system allows for multiple positive steady states, a phenomenon also known as {\em multistationarity}. More precisely, from the stoichiometry and the rate functions one can define a {\em species formation rate function} $F$ having $n$ variables and $n$ values, where $n$ is the number of species. The dynamic of the system consists of the finite set of differential equations
$$
\frac{dc(t)}{dt} = F (c(t)) \, , \leqno{(A)} $$
where $c(t)$ is a vector of concentrations varying with time. We shall consider  a theorem of Wiuf and Feliu on the  injectivity properties of (A) when the rate functions are {\em strictly monotonic} (\cite{WF}, Th. 10.2), a condition fulfilled not only by mass-action kinetics, but also by Michaelis-Menten and Hill kinetics, and by others.

\smallskip

Our work stems from results in the biological context, namely genetic networks theory. This theory models the interactions (activation and repression) between genes in a given cell: if $c(t)$ denotes the vector of concentrations of proteins in this cell at time $t$, one assumes that $c(t)$ obeys an equation like (A). Again, one problem is to give necessary conditions for (A) to allow for multistationarity. A well known conjecture of Thomas \cite{T} (1981), states that such a necessary condition is the fact that the {\em interaction graph} of the jacobian of $F$ contains a positive circuit. Mathematical proofs of this fact were found by Plahte et al. \cite{P}, Snoussi \cite{SN}, Gouz\'e \cite{G}, Cinquin and Demongeot \cite{CD}, and Soul\'e \cite{S}.

\smallskip

In \cite{KST}, another result on gene networks was proved. It says that multistationarity of (A) requires the existence of {\em nuclei} of given signs in the interaction graph of $F$, where a nucleus is a disjoint union of circuits which contains all vertices.

\smallskip
An important step occurred with the paper \cite{SO} by Soliman, which proves a result ``\`a la Thomas'' for chemical reaction networks,  and uses in a non-trivial way the stoichiometric matrix. In fact, as Soliman explained to us, one can recover the system (A) for any function $F$ as the dynamic of a {\em special} chemical reaction network (with ad hoc stoichiometry, see 2.2 below), so that the original Thomas' rule becomes a special case of \cite{SO}.

\smallskip

In this paper we are looking for a common generalization of \cite{KST} and \cite{SO}. We state this as a conjecture about chemical reaction networks (Conjecture~1).  Feliu and Wiuf proved this conjecture  when the rate functions are strictly monotonic (Theorem~3).

\smallskip

Note that an important invariant of a system like (A) is the determinant of the Jacobian matrix of $F$. But, if the system (A) obeys {\em conservation laws}, this determinant is identically trivial. In \cite{WF}, Wiuf and Feliu replace this invariant by the determinant of another matrix, and they give a formula for it as the sum of principal minors of the Jacobian (Section 2, Prop. 1). By solving the conservation laws we derive another formula (Prop. 2).

\smallskip

In Section 3, following \cite{SO}, \cite{GF} and \cite{T}, we introduce the {\em reaction labelled influence graph} of a chemical reaction network, and we use it to formulate the main result of \cite{SO}, with three important corollaries. Next, we state the main result of \cite{KST} (Theorem~2). In Section 4 we present our conjecture (Conjecture 1) and, after defining strictly monotonic kinetics, we state the theorem of Feliu and Wiuf (Theorem 3).

\smallskip

Section 5 is devoted to concrete examples which illustrate
several aspects of the theory. In particular, we use Theorem 3 to modify a reaction network in order to produce bistability. In the last example we consider a case of non-monotone kinetics.

\smallskip

After that, Section 6 (resp. Section 7) gives a proof of Proposition 2 (resp. Theorem 3). Finally, in Section 8, we present some conclusions and remarks.

\smallskip

\bigskip

\section{Conservation laws}

\subsection{Definitions}

Let ${\mathbb R}_+$  be the set of positive  real numbers, and ${\mathbb N}$ the set of natural integers. Let ${\mathcal N} = ({\mathcal S}, {\mathcal C}, {\mathcal R})$ be a {\em chemical network}. As explained in \cite{WF}, Def.~3.2, ${\mathcal N}$ consists in three finite sets:
\begin{enumerate}
\item[$\bullet$] A finite set of {\em species} ${\mathcal S} = \{S_1 , \ldots , S_n\}$;
\item[$\bullet$] A set of {\em complexes} ${\mathcal C} \subset {\mathbb N}^n$;
\item[$\bullet$] A set of {\em reactions} ${\mathcal R} = \{r_1 , \ldots , r_m\} \subset {\mathcal C} \times  {\mathcal C}$.
\end{enumerate}
It is assumed that if $y \in {\mathcal C}$, then $(y,y) \notin {\mathcal R}$ and there exists $y' \in {\mathcal C}$ such that $(y',y) \in {\mathcal R}$ or $(y,y') \in {\mathcal R}$. We write
$$
r : y \longrightarrow y'
$$
to mean that $r = (y,y') \in {\mathcal R}$.

\smallskip

We consider {\em differential kinetics} $K = (K_j)$ for the network ${\mathcal N}$ i.e. (see \cite{WF} Def.~3.3) to each reaction $r_j : y_j \to y'_j$ in ${\mathcal R}$ we attach a {\em rate function}
$$
K_j : {\mathbb R}_+^n \longrightarrow {\mathbb R}
$$
which is differentiable.  Let $A$ be the {\em stoichiometric matrix}, i.e. the $n \times m$ matrix with $j$-th column $y'_j - y_j$, where $(y_j , y'_j)$ lies in ${\mathcal R}$. The {\em species formation rate function} is the map
$$
F : {\mathbb R}_+^n \longrightarrow {\mathbb R}^n
$$
defined by the formula
\be
\label{eq10}
F(c) = \sum_{j=1}^m K_j(c)(y'_j - y_j) \, .
\ee
If $c(t) \in {\mathbb R}_+^n$  is a differentiable function of $t \in {\mathbb R}_+$, and $\dot c (t)$ its derivative, the dynamic of ${\mathcal N}$ consists of the finite set of $n$ ordinary differential equations

$$
\dot c = F(c) \, .
$$

\bigskip

\subsection{A special case}

Let 
$$
F : {\mathbb R}_+^n \longrightarrow {\mathbb R}^n
$$
be any differential map. One can define as follows a chemical network
such that $F$ is the corresponding species formation rate function.
One takes $m=n$, and one defines $y_j$ (resp. $y'_j$)  to be zero
(resp. the complex $y'_{ij}$ such that  $y'_{jj}=1$, and
$y'_{ij}= 0$ when $i \neq j$). The rate function $K_j$ is defined to be
the $j$-th component $F_j$ of the function $F$. The equation 
 (\ref{eq10})  is then clearly satisfied.

\subsection{A formula of Wiuf and Feliu}

We equip ${\mathbb R}^n$ with the standard scalar product. We denote by ${\rm Im} (A)$ the image of $A$ and by ${\rm Im} (A)^{\perp}$ the orthogonal complement of ${\rm Im} (A)$. Assume ${\rm Im} (A)^{\perp}$ has dimension $d $. Consider a {\em reduced basis} of ${\rm Im} (A)^{\perp}$ i.e. a basis $(\omega^1 , \ldots , \omega^d)$ such that $\omega_i^i = 1$ if $1 \leq i \leq d$, and $\omega_j^i = 0$ when $j = 1 , \ldots , \widehat i , \ldots , d$ (such a basis exists after reordering the species, cf. \cite{WF}, Def.~5.1). The formula
$$
\widetilde F(c) = (\omega^1 \cdot c , \ldots , \omega^d \cdot c , F_{d+1} (c) , \ldots , F_n (c))
$$
defines a new map
$$
\widetilde F : {\mathbb R}_+^n \longrightarrow {\mathbb R}^n \, .
$$

When $c \in {\mathbb R}_+^n$ we let $J_c (F)$ (resp. $J_c (\widetilde F)$) be the Jacobian matrix of $F$ (resp. $\widetilde F$) at $c$.

If $I \subset \{ 1 , \ldots , n \}$ we let $J_c (F)_{I,I}$ be the submatrix of $J_c (F)$ with entries having indices in $I \times I$.
The following is proved in \cite{WF}, Prop.~5.3:

\bigskip

\noindent {\bf Proposition 1:}

\begin{equation}
\label{eq1}
\det (J_c (\widetilde F)) = \sum_{I \subset \{ 1 , \ldots , n\} \atop \# \, I = n-d} \det (J_c (F)_{I,I}) \, .
\end{equation}

\medskip

 We simplify the notation by defining
$$
d(c,K) = \det (J_c (\widetilde F)) \, .
$$

\subsection{Another formula}

Let $k_1 , \ldots , k_d$ be $d$ real positive constants. The {\em conservation laws}
\be
\label{eq4}
\omega^i \cdot c = k_i \, , \qquad 1 \leq i \leq d \, ,
\ee
reduce by $d$ the number of variables. Namely, by solving (3), if $1 \leq i \leq d$, we get
$$
c_i = k_i + u_i (c_{d+1} , \ldots , c_n) \, ,
$$
where $u_i (c_{d+1} , \ldots , c_n)$ is a linear combination of the variables $c_{d+1} , \ldots , c_n$. When $d+1 \leq i \leq n$ we define
$$
\varphi_i (c_{d+1} , \ldots , c_n) = F_i (k_1 + u_1 (c_{d+1} , \ldots , c_n) , \ldots , k_d + u_d (c_{d+1} , \ldots , c_n) , c_{d+1} , \ldots , c_n) \, .
$$
to be the function obtained from $F_{i} (c)$ by replacing $c_1 , \ldots , c_d$ by their expression. If $c = (c_{d+1} , \ldots , c_n)$ we let 
$$
\Phi (c) = \left( \frac{\partial \varphi_i}{\partial c_j} (c) \right)_{d+1 \leq i,j \leq n}
$$
be the corresponding Jacobian matrix, and $\Phi' (c)$ the matrix obtained from $\Phi (c)$ by replacing $k_i$ by $\omega^i \cdot c$, $i = 1,\ldots , d$.

\bigskip

\noindent {\bf Proposition 2:} 
$$
d(c,K) = \det (\phi' (c)) \, .
$$

\section{Previous results}

\subsection{The reaction labelled influence graph}

We keep the definitions of 2.1 above and we fix $c \in {\mathbb R}_+^n$. The {\em reaction labelled infuence graph} (RLIG) $G(c)=G(c,K)$ of $K$ at $c$ is defined as follows (cf. \cite{SO}). The set of vertices of $G(c)$ is the set ${\mathcal S}$ of species of ${\mathcal N}$. There is a positive (resp. a negative) edge $i \to i'$ in $G(c)$, labelled by the reaction $r_j : y_j \to y'_j$, when $\frac{\partial F_{i'}}{\partial c_i} (c)$ is positive (resp. negative) and $i$ (resp. $i'$) is a component of $y'_j$ (resp. $y_j$).  In the special case considered in 2.2, the RLIG coincides with the interaction graph attached to $F$.

A {\em circuit} in $G(c)$ is a sequence of distinct vertices $i_1 , \ldots , i_k$ such that there is an edge from $i_{\alpha}$ to $i_{\alpha + 1}$ if $1 \leq \alpha \leq k-1$, and from $i_k$ to $i_1$. The {\em sign} of $C$ is the product of the signs of its edges. Several circuits are said {\em disjoint} when they do not share any vertex. A {\em hooping} is a union of one or several disjoint circuits. If $s \geq 1$ is an integer, an $s$-{\em hooping} is a hooping containing $s$ vertices. When $s=n$, an $s$-hooping is called a {\em nucleus}\cite{KST}.
 The {\em sign of a hooping} $H$ is $\varepsilon(H) = (-1)^{p+1}$, where $p$ is the number of positive circuits contained in $H$ (see  \cite{EL}).

\smallskip

Given a hooping $H$, the {\em restriction} of ${\mathcal N}$ to $H$ is the sub-network ${\mathcal N}_{H}$ of ${\mathcal N}$ where reactions $r_j$ not appearing in $H$ are omitted. Note that, given a vertex $i \in {\mathcal S}$, there is at most one edge issued from $i$ in the RLIG of ${\mathcal N}_{H}$. Therefore the stoichiometric matrix of ${\mathcal N}_{H}$ is a square matrix. Let $\Lambda_H$ be its determinant. We introduce the following

\bigskip

\noindent {\bf Definition 1:} A hooping $H$ of $G(c)$ is called {\em admissible} when the stoichiometric matrix of ${\mathcal N}_{H}$ is invertible (i.e. $\Lambda_H \neq 0$) .

\subsection{A result of Soliman}

A zero $c \in {\mathbb R}_+^n$ of the function $F$ is called {\em non-degenerate} when the Jacobian matrix $\left( \frac{\partial F_{i'}}{\partial c_i} (c) \right)_{1\leq i, i' \leq n}$ is invertible.

\bigskip

\noindent {\bf Theorem 1}  \cite{SO}: Assume $F$ has several non-degenerate zeroes. Then there exists $c \in {\mathbb R}_+^n$ such that $G(c)$ has a positive circuit contained in an admissible hooping.

\bigskip

\subsection{} 

Soliman derived three corollaries from Theorem 1:

\medskip

\noindent {\bf Corollary  1} \cite{SO}: A necessary condition for the multistationarity of a biochemical system is that there exists a positive circuit in its RLIG using at most once each reaction.

\medskip

Indeed, if a circuit uses more than once some  reaction, it is not admissible.

\bigskip

\noindent {\bf Corollary  2} \cite{SO}: A necessary condition for the multistationarity of a biochemical system is that there exists a positive circuit  in its RLIG not using both forward and backward directions  of any reversible reaction.

\medskip

Indeed, if a circuit uses both forward and backward directions  of some reversible reaction, it is not admissible.

And, by a similar argument:

\medskip

\noindent {\bf Corollary  3} \cite{SO}: A necessary condition for the multistationarity of a biochemical system is that there exists a positive circuit in its RLIG not using  all species involved in a conservation law.

\bigskip

\subsection{A result of Kaufman, Soul\'e and Thomas}

In this section we assume that we are in the special case considered in 2.2. Consider the finite set ${\mathcal F}$ of real functions on ${\mathbb R}_+^n$ of the form $\pm \underset{i \in I}{\prod} (\partial F_i / \partial \, x_{\tau (i)})$ where $I \subset \{ 1,\ldots , n\}$ is any subset and $\tau : I \to \{i,\ldots , n \}$ is an injective map. We consider the following condition:

\begin{quotation}
\noindent (C) \quad Given two functions $f$ and $g$ in ${\mathcal F}$ such that $f$ is not identically zero and $g$ is strictly positive somewhere in ${\mathbb R}_+^n$, there exists $x \in {\mathbb R}_+^n$ such that $f(x) \ne 0$ and $g(x) > 0$.
\end{quotation}

\noindent Note that (C) is very often fulfilled.

\smallskip

We say that the interaction graph $G$ has a {\em variable nucleus} when there exist two points $c$ and $d$ in ${\mathbb R}_+^n$ and a nucleus of opposite signs in $G(c)$ and $G(d)$ (see \cite{KST} 3.1 for a more precise definition).

\bigskip

\noindent {\bf Theorem 2} \cite{KST}: Let $F$ be such that (C)  is satisfied. If $F$ has several non-degenerate zeroes, then:
\begin{enumerate}
\item[i)] either there exists $c \in {\mathbb R}_+^n$ such that $G(c)$ has two nuclei of opposite signs;
\item[ii)] or $G$ has a variable nucleus.
\end{enumerate}

\medskip

This theorem is presented as a conjecture in  \cite{TK}.

\section{The main result}

\subsection{A conjecture}

Assume $m \geq 1$ is any positive integer. In view of Theorems 1 and 2 it seems reasonable to make the following conjecture:

\bigskip

\noindent {\bf Conjecture 1:} Assume that $F$ has several non-degenerate zeroes and that (C) is satisfied. If $F$ has $d$ conservation laws, and $s=n-d$,
\begin{enumerate}
\item[i)] either there exists $c \in {\mathbb R}_+^n$ such that $G(c)$ has two admissible $s$-hoopings of opposite signs;
\item[ii)] or $G$ has a variable admissible $s$-hooping.
\end{enumerate}

\subsection{Monotonic kinetics}
The Conjecture 1 is known   when
the kinetics are strictly monotone. To describe
this result, we introduce more definitions from \cite{WF}.

\smallskip

Let $Z = (Z_{ji})$ be a matrix of type $m \times n$ such that each entry $Z_{ji}$ is equal to $+1,-1$ or $0$. We let
$$
Z_j^+ = \{ i  \, {\rm such \,   that}\,  Z_{ji} = + 1 \} \, ,
$$
$$
Z_j^- = \{ i  \, {\rm such \,   that}\,  Z_{ji} = - 1 \} \, ,
$$
and
$$
Z_j^0 = \{ i   \, {\rm such \,   that}\,  Z_{ji} = 0 \} \, .
$$

 We assume that, for every $c \in {\mathbb R}_+^n$ and every $j \in \{1,\ldots ,m\}$, $K_j (c) > 0$.
We say that $K$ is {\em strictly monotonic} with respect to $Z$ (\cite{WF}, Def.~9.3) if, for all $j \in \{1,\ldots ,m\}$, the function $K_j$  is
\begin{enumerate}
\item[i)] strictly increasing in $c_i$ when $i \in Z_j^+$,
\item[ii)] strictly decreasing in $c_i$ when $i \in Z_j^-$,
\item[iii)] constant in $c_i$ when $i \in Z_j^0$.
\end{enumerate}

We denote by $K(Z)$ the set of kinetics $K$ which are strictly monotonic with respect to $Z$.

\smallskip

Let $A$ be the stoichiometric matrix. The matrix $A$ is called {\em not $Z$-injective} over $K(Z)$ when there exist $K \in K(Z)$, $a \in {\mathbb R}_+^n$ and $b \in {\mathbb R}_+^n$ such that $F(a) = F(b)$.

\smallskip

The network ${\mathcal N}$ is called {\em non-degenerate} when there exists $c \in {\mathbb R}_+^n$ and $K \in K(Z)$ such that $d(c,K) \ne 0$.

\medskip

The following result relates the fact that $F$ is not injective to the existence of two admissible $s$-hoopings
of opposite signs in the RLIG. It proves Conjecture 1 when $K$  is strictly monotone: there are no variable $s$-hoopings and,
if F has several non degenerates zeroes, A is not Z-injective
over K(Z) (we do not need to assume that hypothesis (C) is satisfied).

\bigskip

\noindent {\bf Theorem 3} \cite{WF,WF1}

Assume that ${\mathcal N}$ is non-degenerate and that $A$ is not $Z$-injective over $K(Z)$. Let $s=n-d$ be the rank of $A$. Then, for every $c \in {\mathbb R}_+^n$ and $K \in K(Z)$, the RLIG $G(c)$ contains two admissible $s$-hoopings of opposite signs.

\section{Examples}

To illustrate Conjecture 1 we consider four simple examples.
For the first three examples we simply use mass-action kinetics to describe the reaction rates.
The last example involves a biochemical reaction
with non-monotone kinetics.

\bigskip

\noindent {\bf Example 1. An autocatlytic reaction}

\smallskip

Our first example consists of the simple autocatalytic reaction system shown in Fig. 1. It is described by the differential equations
\begin{eqnarray}
\frac{dA}{dt} &= &S_A -k_1 AC^2 + k_{-1} C^3 - \gamma_a A  \\
\frac{dB}{dt} &= &-k_{-2} B + k_{2} C - \gamma_b B \nonumber \\
\frac{dC}{dt} &= &S_C + k_1 AC^2 - k_{-1} C^3 - k_2 C + k_{-2} B - \gamma_c C \nonumber 
\end{eqnarray}
where we consider a constant source $S_A$ of $A$ and $S_C$ of $C$. The associated Jacobian matrix is 
\begin{equation}
J = \left[ \begin{matrix}
-k_1 C^2 - \gamma_a & 0 & -2 k_1 AC + 3 k_{-1} C^2 \\
{ \ } \\
0 & -k_{-2} - \gamma_b & k_2 \\
{ \ } \\
k_1 C^2 & k_{-2} & 2 k_1 AC - 3 k_{-1} C^2 - k_2 - \gamma_c
\end{matrix} \right]  
\end{equation}
which contains the contributions, if any, of each reaction for each of the variables. The corresponding reaction labelled influence graph (RLIG) is depicted in Fig. 2. 

\smallskip

In this RLIG, there are two positive circuits using both the forward and backward directions of a reversible reaction. According to Definition 1 and corollary 2 of Theorem 1 they cannot form admissible hoopings because their corresponding stoichiometric matrices are singular. There remain two admissible positive nuclei formed by the positive loop $R_1$ on $C$, the negative loop $R_a$ on $A$ and, respectively, the negative loop $R_b$ or $R_{-2}$ on $B$. The RLIG also comprises several admissible negative nuclei composed of negative self-loops. The necessary conditions of Theorem 1 and Theorem 3 are satisfied and for suitable parameter values this system exhibits bistability (Fig. 3).

\smallskip

According to (20) below, the determinant expansion of the Jacobian matrix (5) is a sum of terms indexed over all the admissible nuclei
\begin{eqnarray}
\det (J) = & 2 k_1 AC (k_{-2} + \gamma_b) \gamma_a -  3k_{-1} C^2 (k_{-2} + \gamma_b) \gamma_a \nonumber \\
           & - k_2 \gamma_b (\gamma_a + k_1 C^2) -(k_{-2}+\gamma_b)(\gamma_a + k_1 C^2) \gamma_c 
\end{eqnarray}
where the terms associated with the admissible positive and negative nuclei are of opposite sign.

\smallskip

If $\gamma_a = 0$, Theorem 1 does not rule out the possibility of several steady states as there still exists a positive circuit $R_1$ included in two admissible 2-hoopings $R_1 R_b$ and $R_1 R_{-2}$. However, in this case, Theorem 3 is no longer satisfied. This can be seen on the RLIG in Fig. 2 and on the determinant expansion (6) which, for $\gamma_a = 0$, contain only negative admissible nuclei composed of negative self-loops. The system cannot have multiple steady states and its unique steady state can be determined analytically.

\bigskip

\noindent {\bf Example 2. Conservation laws}

\smallskip

To illustrate Theorem 3 in the presence of a conservation law, we consider a model of reversible substrate inhibition studied by Mincheva and Roussel \cite{M}. This biochemical reaction system and corresponding RLIG are shown in Figs. 41 and 42.

\smallskip
As can be seen on Fig. 42, the 3-circuit between $S_1$, $ES_1$ and $E$, involving the reactions $R_5$, $R_4$ and $R_3$, is the only positive circuit that satisfies the necessary conditions of the three corollaries of Theorem 1. It forms an admissible positive 4-hooping with the negative loop $R_7$ on $P$.

\smallskip

The system of differential equations is
\begin{eqnarray}
\dfrac{dx_1}{dt} &=& -k_1x_1+k_2-k_3x_1x_2-k_5x_1x_3+k_6x_4  \\
\dfrac{dx_2}{dt} &=& -k_3x_1x_2+k_4x_3 \nonumber \\
\dfrac{dx_3}{dt} &=& k_3x_1x_2-k_4x_3-k_5x_1x_3+k_6x_4 \nonumber \\
\dfrac{dx_4}{dt} &=& k_5x_1x_3-k_6x_4 \nonumber \\
\dfrac{dx_5}{dt} &=& k_4x_3-k_7x_5 \nonumber
\end{eqnarray}
where $x_i$, $i =1, \ldots 5$, are the concentrations of species $S_1$, $E$, $ES_1$, $S_1$ $ES_1$ and $P$, respectively. The enzymatic species are linked by the conservation relation
\begin{equation}
x_2+x_3+x_4=e_t 
\end{equation}
where $e_t$ is the total enzyme concentration.

\smallskip

Due to the conservation relation, the Jacobian matrix of system (7) is singular. Therefore, following the theory developed in the sections 2 and 6 (see also \cite{FW}, 3.1), we construct a $5 \times 5$ matrix $M$ by replacing the fourth row of the Jacobian matrix of eqs (7) by the vector $w^1 = (0, 1, 1, 1, 0)$ defined by the conservation law (3) with $c = x$ and $k_1 = e_t$
\begin{equation}
M = \left[ \begin{matrix}
-k_1-k_3x_2-k_5x_3 & -k_3x_1 & -k_5x_1 & k_6 & 0 \\[7pt]
-k_3x_2 & -k_3x_1 & k_4 & 0 & 0 \\[7pt]
k_3x_2-k_5x_3 & k_3x_1 & -k_4-k_5x_1 & k_6 & 0 \\[7pt]
0 & 1 & 1 & 1 & 0 \\[7pt]
0 & 0 & k_4 & 0 & -k_7 \\
\end{matrix} \right]  
\end{equation}
and
\begin{equation}
\begin{aligned}
\det(M) = & k_1k_4k_6k_7 + k_1k_3x_1k_6k_7 + k_1k_3x_1k_5x_1k_7   \\ 
          & +k_3x_2k_4k_6k_7 - k_3x_1k_4k_5x_3k_7  \, .
\end{aligned}
\end{equation}

The determinant of $M$ contains a sum of terms associated with the positive admissible $s$-hooping (with $s = n-d = 4$) and with the four admissible negative $s$-hoopings which are all formed by a combination of self-loops. The necessary condition for multistationarity is thus satisfied and for suitable parameter values this system is bistable \cite{M}.

\bigskip

\noindent {\bf Example 3. Synthetic approach}

\smallskip

Let us consider the well-known Brusselator \cite{MK3,MK4,MK8} shown in Fig. 5(a). The initial and final product concentrations $A, B, D, E$ are maintained time-independent. 

\smallskip

This model system is described by the differential equations
\begin{eqnarray}
\frac{dX}{dt} &= &A -k_1 B X + k_2 X^2 Y - k_3 X  \\
\frac{dY}{dt} &= &k_1 B X - k_2 X^2 Y \nonumber
\end{eqnarray}
with the associated Jacobian matrix
\begin{equation}
J = \left[ \begin{matrix}
-k_1 B + 2 k_2 X Y - k_3 & k_2 X^2  \\
{ \ } \\
k_1 B - 2 k_2 X Y & -k_2 X^2  
\end{matrix} \right] \, . 
\end{equation}
Equations (11) admit a unique steady state solution
\begin{equation}
X_S=\dfrac{A}{k_3} \quad \quad \quad Y_S=\dfrac{k_1k_3B}{k_2A} 
\end{equation}
that becomes unstable through a Hopf bifurcation leading to sustained oscillations (Fig. 7A).

\smallskip

From the corresponding RLIG (Fig. 5(b)) or Jacobian matrix (12) it can be seen that the Brusselator, in its original form, does not satisfy the conditions of Theorem 3 and Theorem 1: the positive loop $R_2$ of $X$ on itself and the positive circuit between $X$ and $Y$, involving reactions $R_1$ and $R_2$, are not admissible. The determinant of Jacobian (12) contains only one, negative nucleus
\begin{equation}
\det(J) = (-k_2 X^2)  (-k_3) \, . 
\end{equation}

However, inspection of the RLIG in Fig. 5(b) suggests that a slight modification of this influence graph consisting in the addition of a linear decay reaction for species $Y$ (Figs. 6(a) and (b)) allows to satisfy the conditions of both theorems. In addition to three terms corresponding to negative nuclei, the determinant expansion of the modified Jacobian now contains a term associated with a positive nucleus formed by the positive loop $R_2$ on $X$ and the negative loop $R_4$ on $Y$
\begin{equation}
\det(J) = k_3 k_4 + k_3 k_2 X^2 + k_4 k_1 B - 2k_4 k_2 XY 
\end{equation}
where $k_4$ is the kinetic constant of reaction $R_4$.

\smallskip

The steady state solutions are now given by a cubic equation and as shown in Fig. 7B, for suitable parameter values, this slightly modified Brusselator becomes a bistable system without losing the capacity to exhibit oscillations (Fig. 7D). Excitability linked to multistationarity is also observed (Fig. 7C). In Fig. 7D one can see that $k_4$ favors bistability while $k_3$ favors sustained oscillations.

\bigskip

\noindent {\bf Example 4. A variable nucleus}

\smallskip

To illustrate point ii) of conjecture 1, we consider a substrate cycle model derived from the system studied in detail by Hervagault and Canu \cite{MK5,MK6}. In this system two metabolites $S_1$ and $S_2$ are interconverted by two antagonist enzymes. Metabolite $S_1$ is converted into metabolite $S_2$ by enzyme $E_1$ and $S_2$ is converted in turn into $S_1$ by enzyme $E_2$. In addition, enzyme $E_1$ is inhibited by excess of its substrate $S_1$. This cyclic pathway is depicted in Fig. 8 together with the corresponding RLIG. It is governed by the differential equations
\begin{eqnarray}
\dfrac{dS_1}{dt} &=& v_s - \dfrac{v_1 S_1}{K_1+S_1+kS_1^2} + \dfrac{v_2 S_2}{K_2+S_2}-d_1 S_1  \\
\dfrac{dS_2}{dt} &=& \dfrac{v_1 S_1}{K_1+S_1+kS_1^2} - \dfrac{v_2 S_2}{K_2+S_2}-d_2S_2 \nonumber 
\end{eqnarray}
where we consider a constant source $v_s$ of $S_1$ and linear decays of $S_1$ and $S_2$. In this example, the reaction rate of the first reaction is a non-monotone function of its argument (see insert in Fig. 9) and does not satisfy the premises of Theorem 3. Note that this function is a variant
of Example 2.

\smallskip

The RLIG contains a sign-variable self-loop $R_1$ on $S_1$ which together with the negative sef-loop $R_4$ on $S_2$, forms a variable nucleus.

\smallskip

From the Jacobian matrix of (16) 
\begin{equation}
J = \left[ \begin{matrix}
-d_1+\dfrac{v_1 (kS_1^2-K_1)}{(K_1+S_1+kS_1^2)^2} & \dfrac{K_2v_2}{(K_2+S_2)^2}  \\[9pt]
\dfrac{v_1 (K_1-kS_1^2)}{(K_1+S_1+kS_1^2)^2} & -d_2-\dfrac{K_2v_2}{(K_2+S_2)^2} \\\end{matrix} \right] 
\end{equation}
one observes that the variable self-loop has a positive effect on $S_1$ when $S_1 > \sqrt{(K_1/k)}$ and it provides the possibility of a change of sign of the determinant expansion of $J$
\begin{equation}
\det(J)=d_1d_2-d_2\dfrac{v_1 (kS_1^2-K_1)}{(K_1+S_1+kS_1^2)^2}+\dfrac{K_2v_2d_1}{(K_2+S_2)^2} \, . 
\end{equation}
As shown in Fig. 9, for appropriate values of the parameters, this model system has several steady states.

\section{Proof of Proposition 2}

To prove Proposition 2  we first notice that, since $(\omega^1 , \ldots , \omega^d)$ is reduced, when $k=1,\ldots , d$, we have
$$
u_k (c_{d+1} , \ldots , c_n) = - \sum_{j=d+1}^n \omega_j^k  c_j \, .
$$
Therefore, if $j = d+1 , \ldots , n$ and $i = d+1 , \ldots , n$, we have
\begin{equation}
\label{eq2}
\frac{\partial \varphi_i}{\partial c_j} (c) = - \sum_{k=1}^d \omega_i^k \frac{\partial F_{i}}{\partial c_k} (c) + \frac{\partial F_{i}}{\partial c_j} (c) \, .
\end{equation}
Consider the $n$ by $n$ matrix $J_c (\tilde F)$. For every $k = 1,\ldots ,d$, substract from the $j$-th column of this matrix, $j=d+1,\ldots ,n$, the product of the $k$-th column by $\omega_j^k$. The determinant does not change by this operation, and the matrix $M$ obtained this way, is such that the first $d$ lines of $M$ are equal to $\delta_{ij}$, $i = 1,\ldots , d$, $j = 1,\ldots , n$.

\smallskip

From (\ref{eq2}) we deduce that
$$
\det (M) = \det \left( \frac{\partial \varphi_i}{\partial c_j} (c) \right)_{d+1 \leq i,j \leq n} \, .
$$
Since $\det (M) = \det (J_c (\widetilde F))$ Proposition~2 follows.

\section{Proof of Theorem 3}

\subsection{ \ }

  Theorem 3 is due to Feliu and Wiuf, see \cite{WF1} (based on \cite{WF},
Section 11). 
The proof can be described as follows. According to \cite{WF}, Theorem 10.2, and the line after it, since $A$ is not injective over $K$, property i) in \cite{WF}, loc. cit., does not hold, i.e. there exist $c_1 \in {\mathbb R}_+^n$ and $K_1 \in K(Z)$ such that $d(c_1,K^1)=0$.

\smallskip

On the other hand, since ${\mathcal N}$ is non degenerate, there exists $c_2 \in {\mathbb R}_+^n $ and $K^2 \in K_d (Z)$ such that $d(c_2,K^2) \ne 0$, say $d(c_2,K^2) > 0$. By (\ref{eq1}) this implies that there exists $I \subset \{ 1 , \ldots , n \}$ with $\# \, I = s$ such that $\det (J_{c_2}(F^2)_{I,I}) > 0$, where $F^2$ is the species formation rate function defined from $K^2$.  Let $H_s(c_2,K^2)$ be the set of s-hoopings in $G(c_2,K^2)$
and  ${\mathcal H}_2 \subset {\mathcal H}_s (c_2 , K^2)$ (resp. ${\mathcal H}_1 \subset {\mathcal H}_s (c_1 , K^1)$)  be the collection of hoopings in $G(c_2 , K^2)$ (resp. $G(c_1 , K^1)$) having $I$ as set of vertices.
We also write ${\mathcal H} =  {\mathcal H}_s (c, K)$.

\subsection{ \ } We shall need some notions and results from \cite{KA} and \cite{SO}. If $s_i \in {\mathcal S}$, $r_j \in {\mathcal R}$, $c \in {\mathbb R}_+^n$ and $K \in K_d (Z)$ we let
$$
\lambda (s_i , r_j) = \frac{\partial K_j}{\partial c_i} (c)
$$
and
$$
\lambda (r_j , s_i) = y'_{ij} - y_{ij} \, .
$$
For a path $P$ in $G(c,K)$ we let $\lambda_{SR} (P)$ (resp. $\lambda_{RS} (P)$) be the product of all the numbers $\lambda (s_i , r_j)$ (resp. $\lambda (r_j , s_i)$) where $S_i \xrightarrow{ \ R_j \ }$ (resp. $\xrightarrow{ \ R_j \ } S_i$) is an edge in $P$. We define also
$$
\lambda (P) = \lambda_{RS} (P) \, \lambda_{SR} (P) \, .
$$

If $H,H' \in {\mathcal H}  =  {\mathcal H}_s (c , K)$  we write that $H \sim H'$ when $H$ and $H'$ have the same species to reactions arcs $S_i \xrightarrow{ \ R_j \ }$. Note that $\lambda_{SR} (H)$ and $\Lambda (H)$ depend only on the equivalence class $[H] \in {\mathcal H} / \sim$. According to  \cite{KA} Thm. 6.7 (see also \cite{SO}, Thm.~2), we have
\begin{equation}
\label{eq3}
\det (J_c (F)) = \sum_{[H] \in {\mathcal H} / \sim} \Lambda (H) \, \lambda_{SR} (H) \, ,
\end{equation}
and
\begin{equation}
\label{eq4}
\Lambda (H) = \sum_{H' \in [H]} \sigma (H') \, \lambda_{RS} (H') \, ,
\end{equation}
where $\sigma (H') = \pm \, 1$. As a consequence  (\cite{KA} Lemma 6.2  and \cite{SO}, (1)) we get
\begin{equation}
\label{eq5}
\det (J_c (F)) = \sum_{H \in {\mathcal H}} \sigma (H) \, \lambda (H) \, .
\end{equation}

\subsection{ \ } Let us come back to the situation of \S7.1.  Let $J = J_{c_2} (F^2)_{I,I}$. Since $\det (J) > 0$, it follows from (\ref{eq3}), (\ref{eq4}) and (\ref{eq5}) that there exists $H_2 \in {\mathcal H}_2$ such that $\Lambda (H_2) \ne 0$ and $\sigma (H_2) \, \lambda (H_2) > 0$.

\smallskip

We claim that there exists also $H_3 \in {\mathcal H}_2$ such that $\Lambda (H_3) \ne 0$ and $\sigma (H_3)$ $\lambda (H_3) < 0$. Suppose, by contradiction, that, for any $H \in {\mathcal H}_2$ such that $\Lambda (H) \ne 0$, we have $\sigma (H) \, \lambda (H) \geq 0$. Since the kinetics are strictly monotonic, for any $H \in {\mathcal H}_s (c,K)$ the sign of $\sigma (H) \, \lambda (H)$ does not depend on $c$ and $K$, but only on the arcs contained in $H$. Therefore, for any $s$-hooping  $H'$ in ${\mathcal H}_1$ such that $\Lambda (H') \ne 0$, one has  $\sigma (H') \, \lambda (H') \geq 0$, and there exists $H_1 \in {\mathcal H}_1$ such that $\sigma (H_1) \, \lambda (H_1) > 0$. By (\ref{eq5}) this implies that $ \det(J_{c_1} (F^1)_{I,I}) > 0$. But $d(c_1,K_1) = 0$, so we conclude by (\ref{eq1}) that there exists $I'$ of cardinality $s$
such that   $\det(J_{c_1} (F^1)_{I',I'}) < 0$. Therefore, when  $H \in {\mathcal H}$  and $ \Lambda (H) \ne 0$, the number  $\sigma (H) \, \lambda (H) $
takes both positive and negative values.

\smallskip

To conclude the proof of Theorem 3, it remains to note that the sign $\varepsilon (H)$ is $(-1)^{s+1}$ times the sign of $\sigma (H) \, \lambda (H)$ (cf. \cite{EL}, Appendix, Lemma 2). Therefore $G(c)$ contains two admissible $s$-hoopings of opposite signs.

\bigskip

\section{\bf Conclusion} In this paper, we conjecture that the multistationarity of a (bio)chemical network
requires the existence, in the reaction labelled influence graph, of two admissible hoopings of maximal rank and opposite signs
or a variable admissible hooping of maximal rank. This conjecture has been proved \cite{WF1} under
the hypothesis that the kinetic is strictly monotonic in the sense of
\cite{WF} . This includes mass-action kinetics, Michaelis-Menten and Hill kinetics, and  many more. The  proof that we present
makes an essential use of the connection between the reaction labelled influence graph and the determinant of the Jacobian
(or its substitute when there exists conservation laws).
We hope that this theorem can be extended to include arbitrary differentiable metrics (as suggested by Example 4).

This  result can be used to rule out the possibility of multistationarity
in situations where other criteria do not (Example 1).
Given a network which does not fulfill our condition for multistationarity,
following Theorem 3,  inspection of the reaction labelled influence graph or Jacobian matrix
 also suggests ways of modifying the network so that it fulfills our condition (Example 3).

This theorem  is thus an illustration of R.Thomas' insight that, in biology,
one should often look for necessary rather than sufficient conditions.

\bigskip

\bigskip

\centerline {\bf Acknowledgements}

\medskip

We dedicate this work to the memory of our friend Ren\'e Thomas. C.S. expresses his admiration for  the scientific trajectory
of R.Thomas, from fundamental biology to pure mathematics.

\bigskip

We are grateful to E. Feliu, D. Gonze and S. Soliman for their help. The bifurcation diagrams were generated with AUTO \cite{MK7}.
This research did not receive  any specific grant from funding agencies in the public, commercial, or not-for-profit sectors.

\vglue 2cm

\vglue 3cm

\noindent {\bf Figure captions}

\bigskip

\noindent Figure 1: An autocatalytic reaction model. A linear decay is considered for each species. Constant sources $S_A$ of $A$ and $S_C$ of $C$ are not shown.

\medskip

\noindent Figure 2: Reaction labelled influence graph corresponding to the reaction network shown in Fig.1. Each arrow is labelled by the reaction that is involved. Normal arrows correspond to a positive influence, blunt arrows to a negative influence.

\medskip

\noindent Figure 3: Steady state concentration $B$ as a function of the autocatalytic kinetic constant $k_1$, calculated from eqs. (4). This bifurcation diagram is obtained for the parameter values: $S_A = 1$, $k_{-1} = 2$, $\gamma_a = 0.25$, $k_2 = 1$, $k_{-2} = 0.5$, $\gamma_b = 0.1$, $S_C = 10^{-4}$, $\gamma_c = 0.3$. It shows an irreversible transition from the upper stable branch down to the lower stable branch, as a function of $k_1$. Unstable steady states are indicated by the dotted line.

\medskip

\noindent Figure 41: An enzymatic reaction network comprising a substrate inhibition step. The total enzyme concentration is conserved.Figure 42: Reaction labelled influence graph corresponding to the reaction network shown in Fig. 41. Each arrow is labelled by the reaction that is involved. Normal arrows correspond to a positive influence, blunt arrows to a negative influence. Note that there is a positive circuit $R_3R_5R_6R_4$ comprising all the species involved in a conservation law.

\medskip

\noindent Figure 5: (a) The Brusselator. (b) Corresponding reaction labelled influence graph. Each arrow is labelled by the reaction that is involved. Normal arrows correspond to a positive influence, blunt arrows to a negative influence. $R_S$ corresponds to a constant source reaction.

\medskip

\noindent Figure 6: (a) The modified Brusselator which includes a linear decay for $Y$. (b) Corresponding reaction labelled influence graph. Each arrow is labelled by the reaction that is involved. Normal arrows correspond to a positive influence, blunt arrows to a negative influence. There is now an additional negative self-loop on $Y$ related to reaction $R_4$.

\medskip

\noindent Figure 7: Dynamical behaviours. (A) Nullclines (in blue for $X$, red for $Y$) and limit cycle (in black) obtained for the parameter values $A = 2$, $B = 6$, $k_1 = k_2 = k_3 = 1$, $k_4 = 0.15$. (B) Bistability as a function of $k_4$ for $A = 0.1$, $B = 0.25$, $k_1 = k_2 = 1$, $k_3 = 0.001$. (C) Excitability for $A = 1$, $B = 6$, $k_1 = k_2 = 1$, $k_3 = 0.4$, $k_4 = 0.1$. (D) Oscillations (blue circles) and bistability (red triangles) as a function of $k_3$ and $k_4$ observed for randomly generated parameter values (1000 samples, values are taken from a uniform distribution in the following ranges: $A = [0,10]$, $B = [0,10]$, $k_3 = [0,1]$, $k_4 = [0,0.25]$). Light grey circles correspond to a single stable steady state.

\medskip

\noindent Figure 8: Substrate cycle with two opposite interconversions, one of which is inhibited by excess of its own substrate. (a) Reaction network. It includes a first-order decay for both substrates. (b) Corresponding reaction labelled influence graph. Each arrow is labelled by the reaction that is involved. Normal arrows correspond to a positive influence, blunt arrows to a negative influence. The $R_1$-loop of of $S_1$ on itself has a dual arrow (normal and blunt) to indicate that reaction $R_1$ can have a negative or positive influence on $S_1$.

\medskip

\noindent Figure 9: Bistability for the model defined by eqs. (16). The insert shows the non-monotone function $F(S_1) = \frac{v_1 S_1}{K_1 + S_1 + kS_1^2}$ as a function of $S_1$, for $v_1 = 1$, $K_1 = 1$ and $k=1$. The bifurcation diagram shows the concentration of $S_1$ as a function of the inhibition parameter $k$. Other parameter values are: $v_S = 0.01$, $v_1 = 1$, $v_2 = 0.3$, $d_1 = d_2 = 0.001$, $K_1 = K_2 = 1$.

\newpage

$$
\includegraphics[width=15cm]{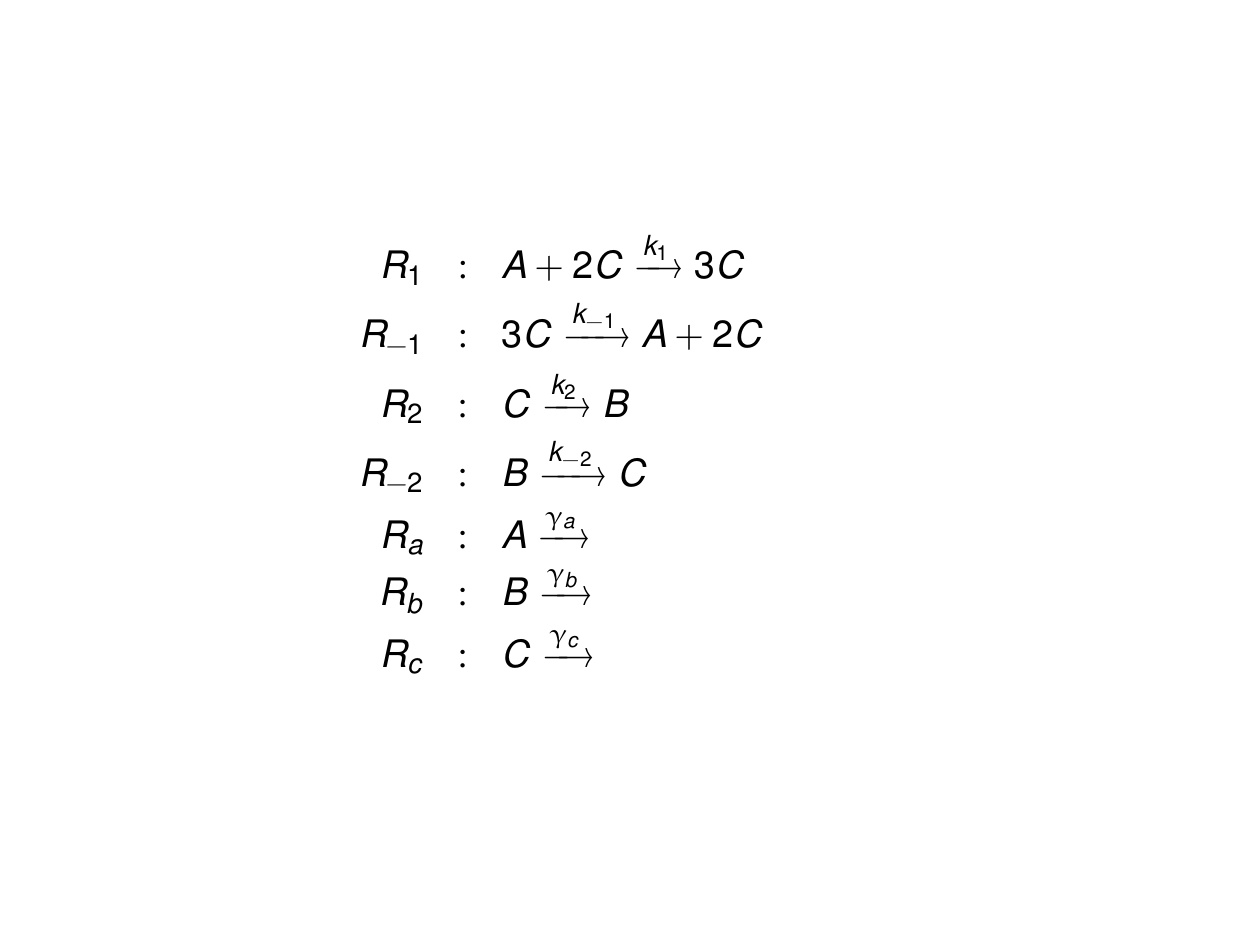}
$$

\centerline {\bf Figure 1}

\newpage

$$
\includegraphics[width=10cm]{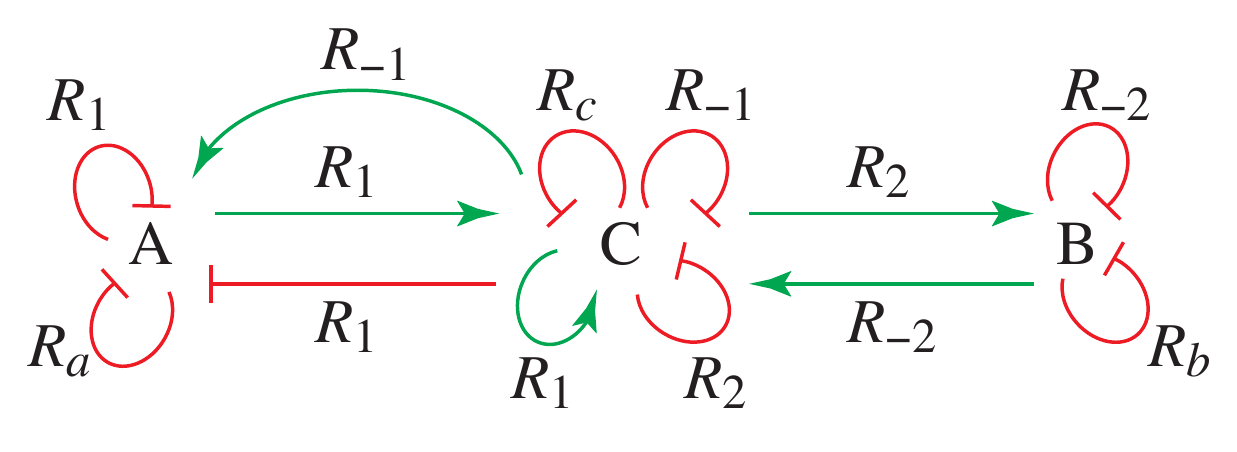}
$$

\centerline {\bf Figure 2}

\newpage

$$
\includegraphics[width=12cm]{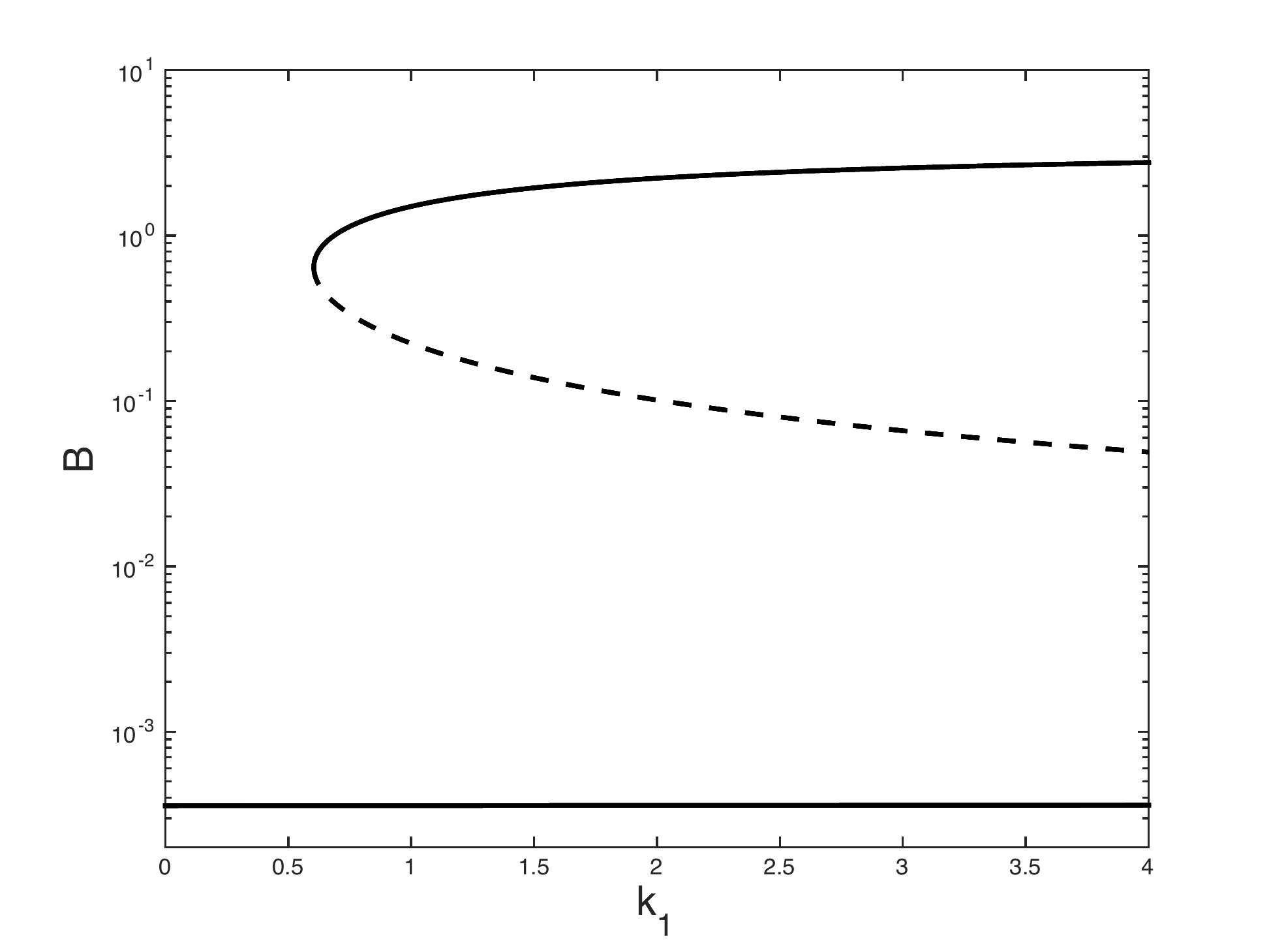}
$$

\centerline {\bf Figure 3}

\newpage

$$
\includegraphics[width=15cm]{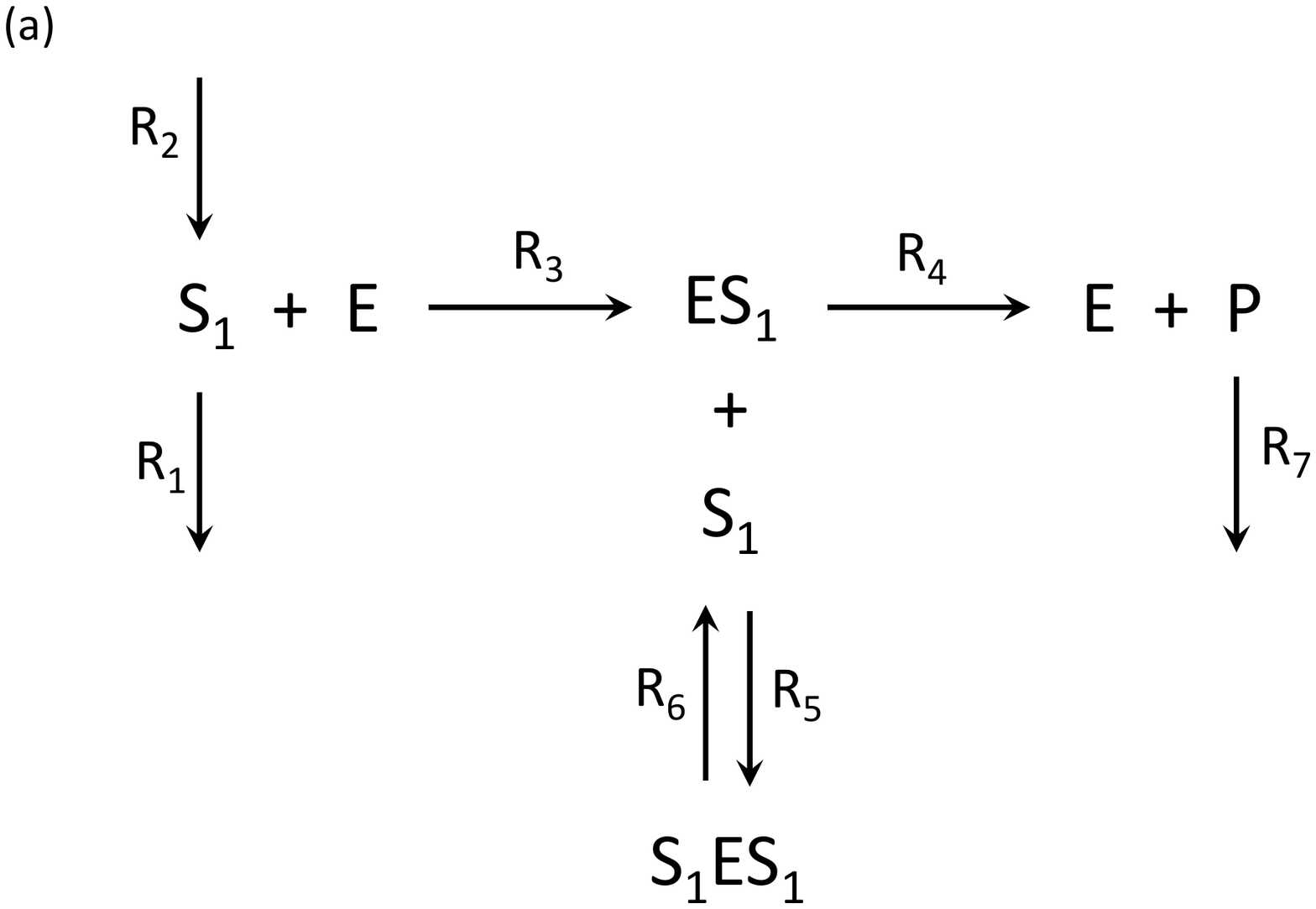}
$$

\centerline {\bf Figure 4a}

\newpage

$$
\includegraphics[width=15cm]{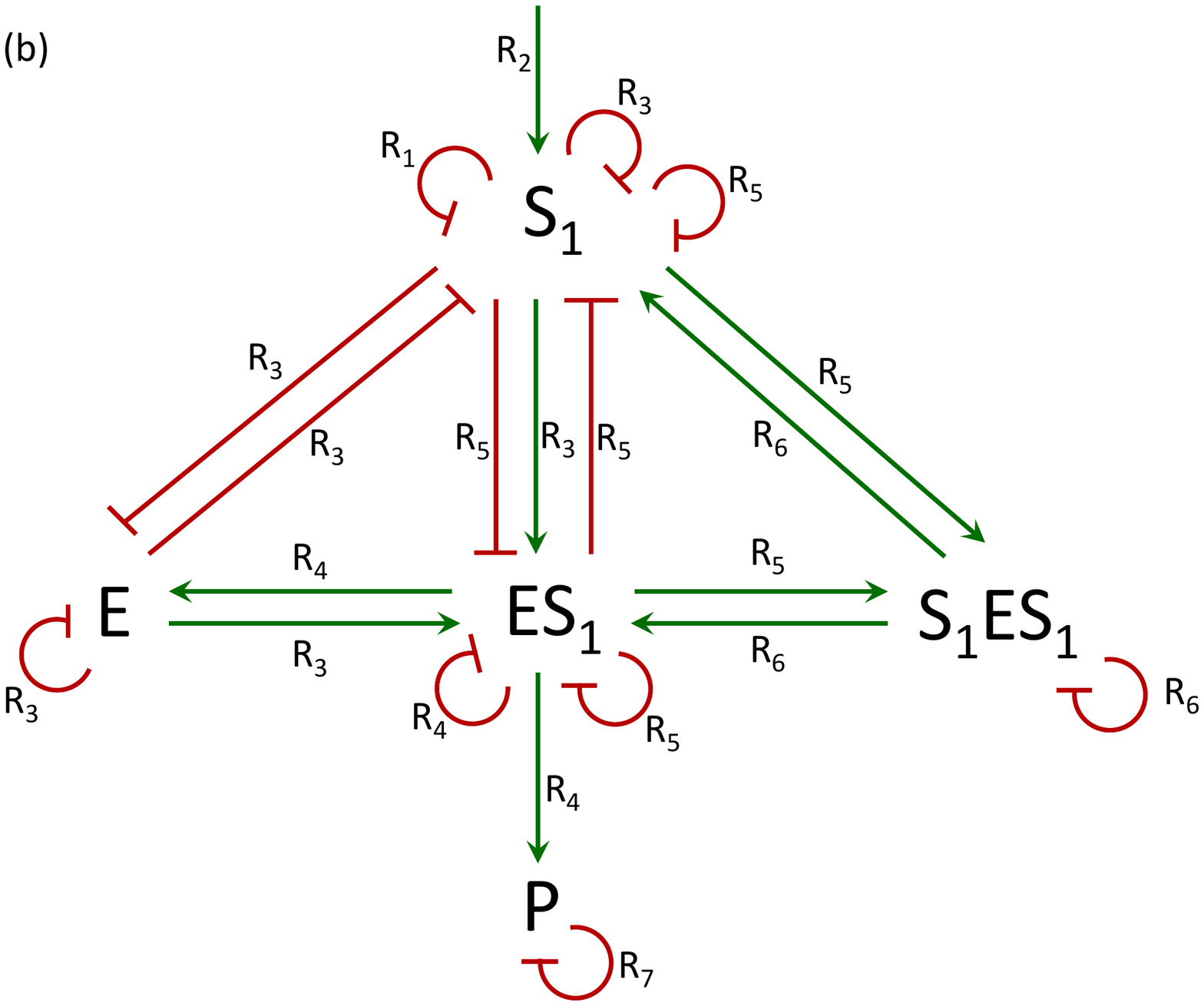}
$$

\centerline {\bf Figure 4b}

\newpage

$$
\includegraphics[width=12cm]{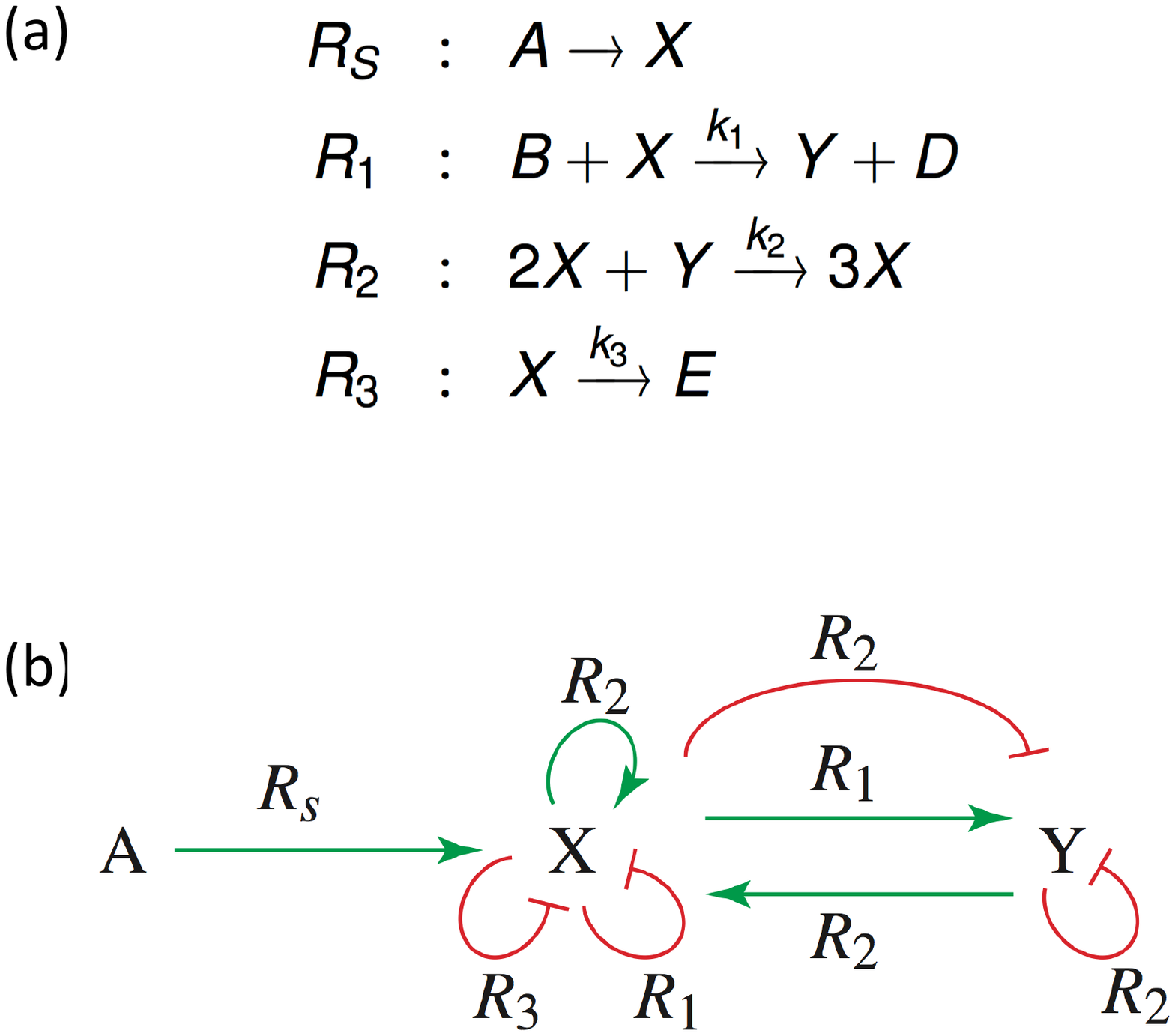}
$$
\centerline {\bf Figure 5}

\newpage

$$
\includegraphics[width=10cm]{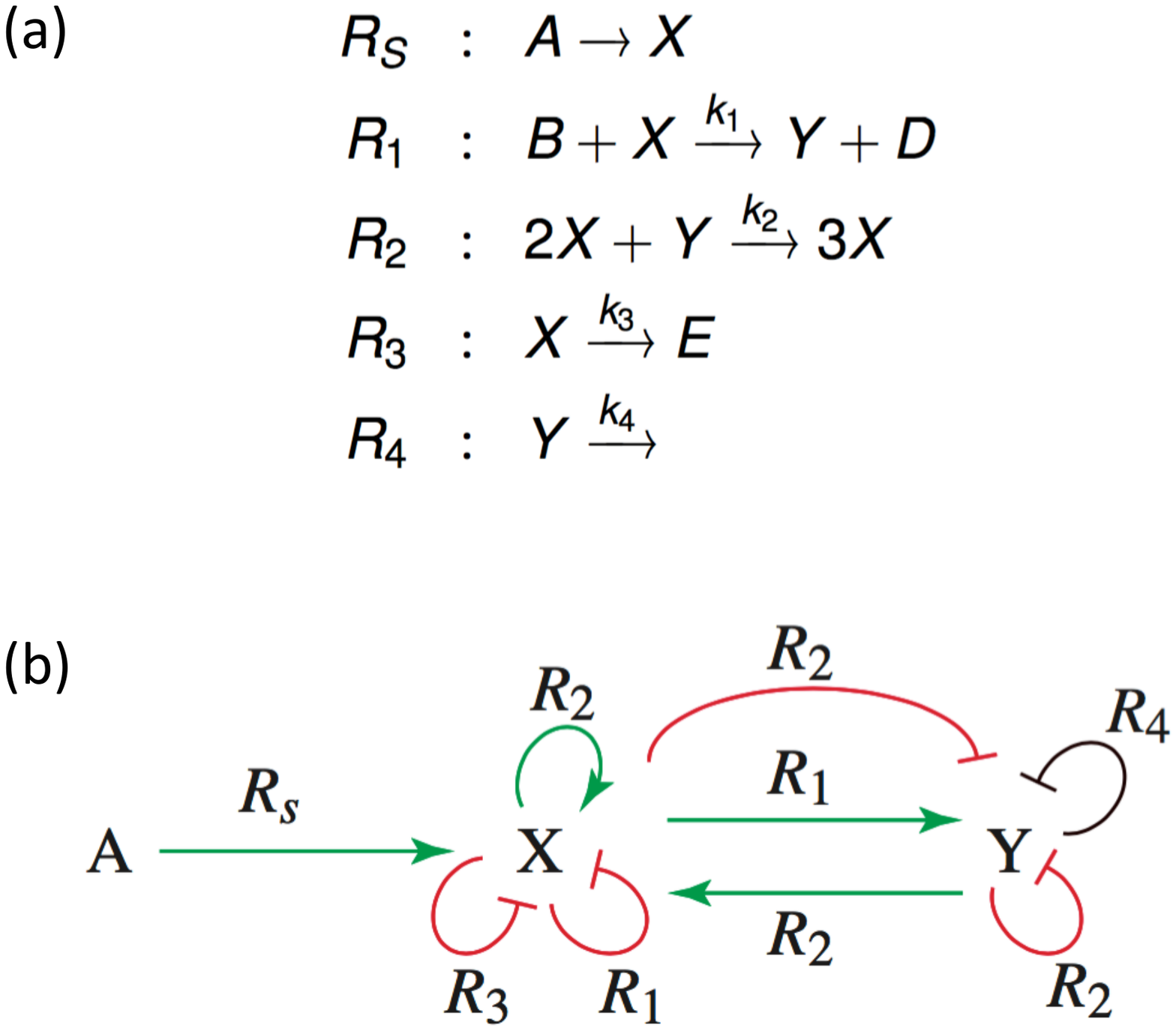}
$$
\centerline {\bf Figure 6}

\newpage

$$
\includegraphics[width=18cm]{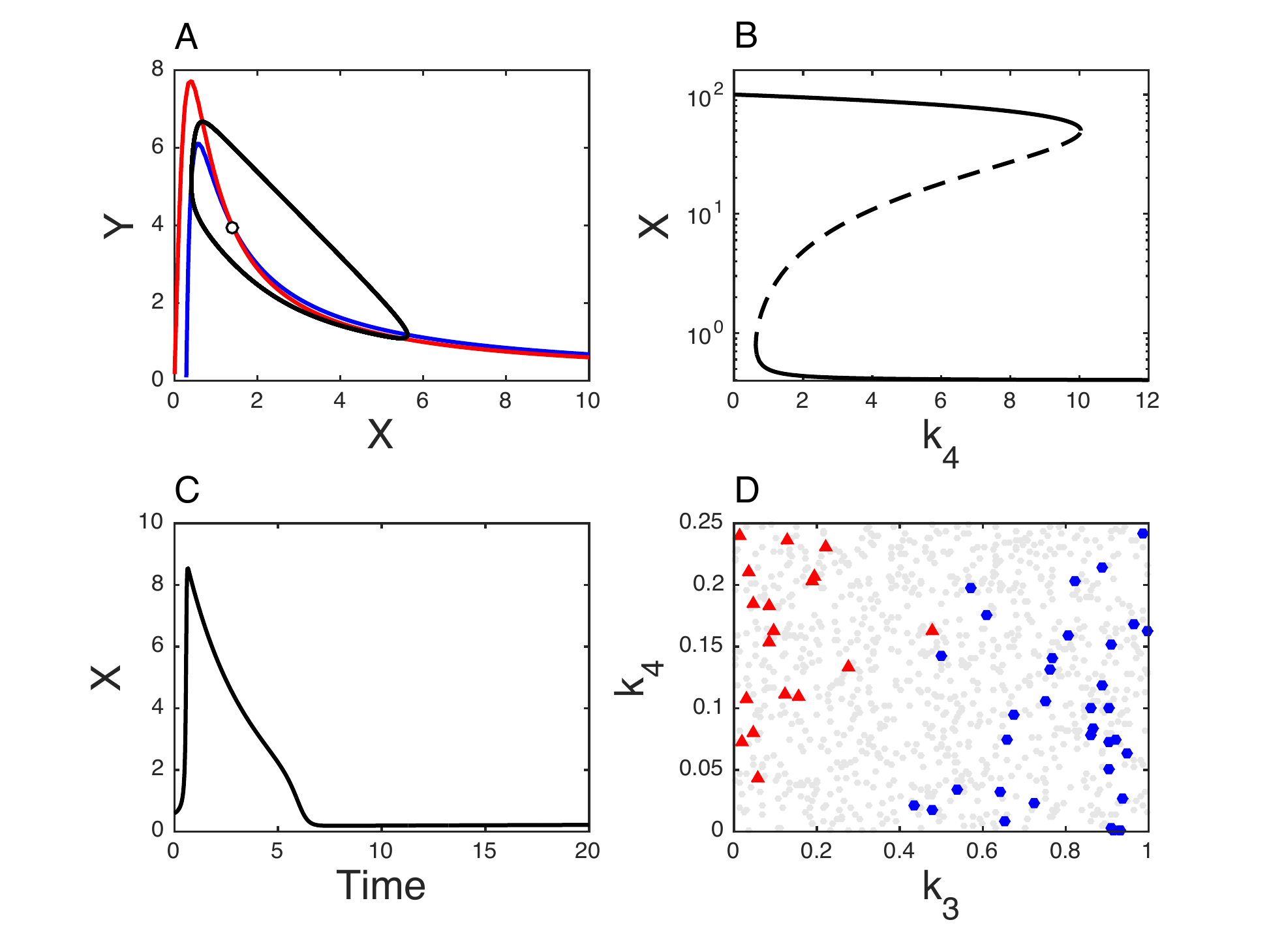}
$$
\centerline {\bf Figure 7}

\newpage

$$
\includegraphics[width=15cm]{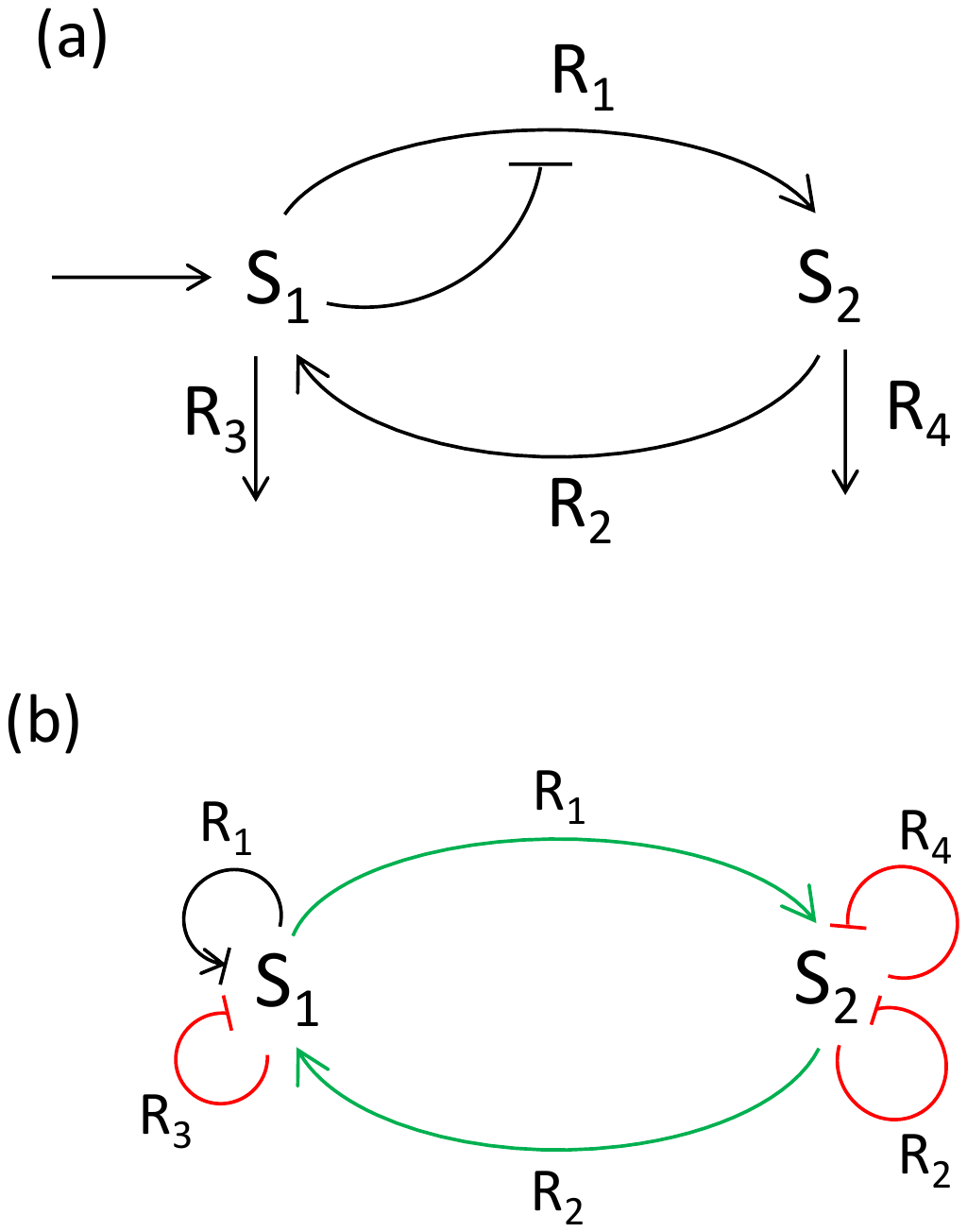}
$$

\centerline {\bf Figure 8}

\newpage

$$
\includegraphics[width=11cm]{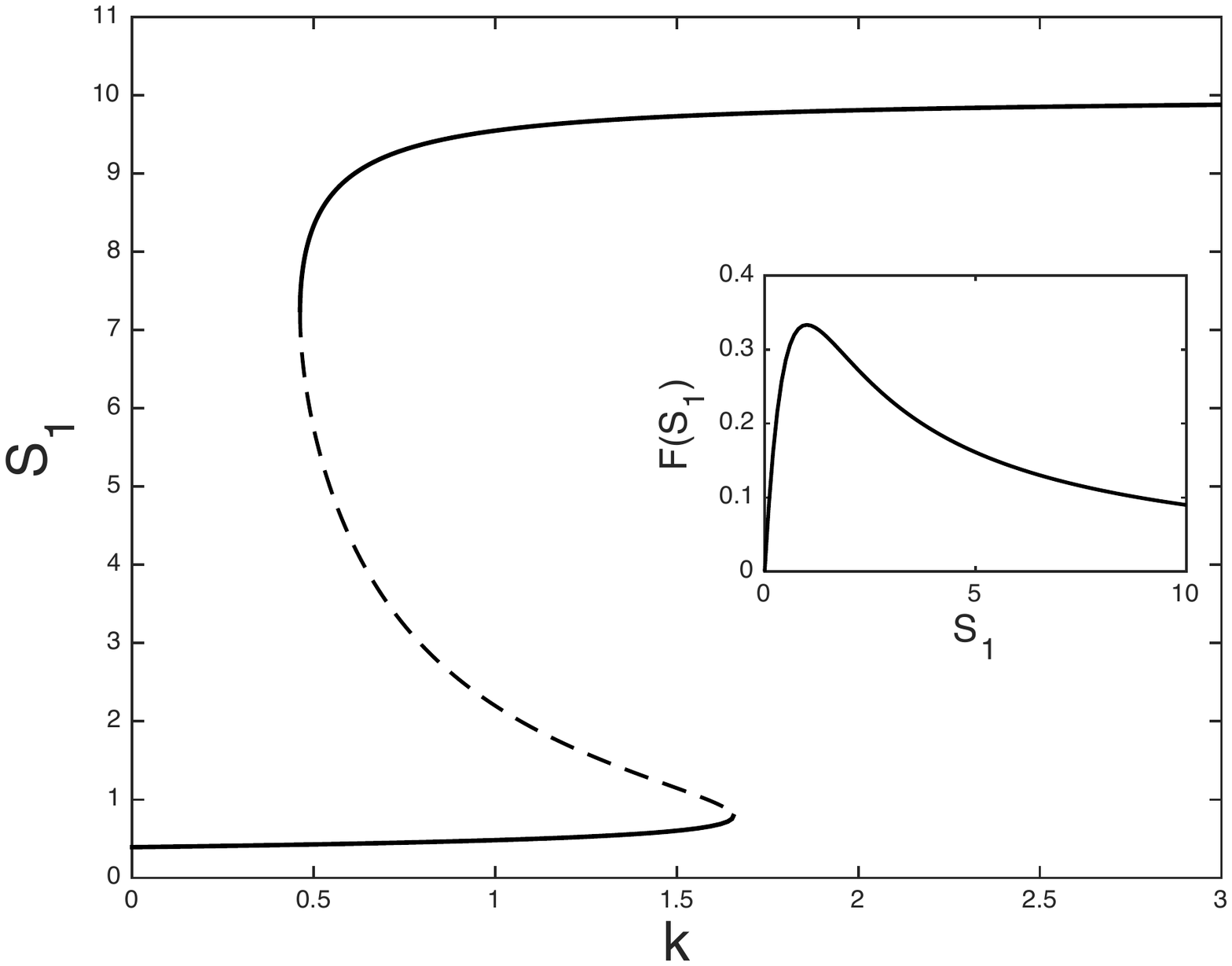}
$$
\centerline {\bf Figure 9}

\end{document}